\documentclass[
  journal=pasa,
  manuscript=research-paper, 
  year=202X,
  volume=YY,
]{cup-journal}

\usepackage{amsmath}
\usepackage{amssymb,microtype,siunitx,booktabs}
\usepackage{amsmath}
\usepackage{amssymb}
\usepackage{multirow}
\usepackage{makecell}
\usepackage{hyperref}
\usepackage[normalem]{ulem}
\usepackage{tabularx}

\newcommand{\Msun}{\,{\rm M}_{\odot}}

\newcommand{\iso}[2]{\hbox{${}^{#1}{\rm #2}$}}

\title{Using Binary Population Synthesis to Calculate the Yields of Low- and Intermediate-Mass Binary Populations at Low Metallicity}

\author{Zara Osborn}
\affiliation{School of Physics \& Astronomy, Monash University, Clayton 3800, VIC, Australia}
\email[Z. Osborn]{zara.osborn@monash.edu}

\author{Amanda I. Karakas}
\affiliation{School of Physics \& Astronomy, Monash University, Clayton 3800, VIC, Australia}

\author{Devika Kamath}
\affiliation{School of Mathematical and Physical Sciences, Macquarie University, Balaclava Road, Sydney, NSW 2109, Australia}

\author{Robert G. Izzard}
\affiliation{Astrophysics Research Group, University of Surrey, Guildford, Surrey GU2 7XH, UK}

\author{Alex J. Kemp}
\affiliation{Institute of Astronomy, KU Leuven, Celestijnenlaan 200D, 3001 Leuven, Belgium}

\author{Chiaki Kobayashi}
\affiliation{Centre for Astrophysics Research, Department of Physics, Astronomy and Mathematics, University of Hertfordshire, Hatfield, AL10 9AB, UK}

\received {dd Mmm YYYY}
\revised  {dd Mmm YYYY}
\accepted {dd Mmm YYYY}
\published{22 September 202X}


\begin{document}

\begin{abstract}
Asymptotic giant branch (AGB) stars are important to chemical evolution at metallicity $Z \sim 0.0001$ ($\text{[Fe/H]} \approx -2.2$) as they contribute significantly to the production of nitrogen, lead, and dust in the early Universe. The contribution of AGB stars to the chemical evolution of the Universe is often quantified using the chemical yields from single AGB stars. Binary evolution challenges our understanding of chemical evolution as binary phenomena such as mergers and mass transfer episodes can significantly alter the stellar evolution pathways and yields. In this work, we use binary population synthesis code \textsc{binary\_c} to model populations of low and intermediate-mass ($0.7-7\Msun$) stars at metallicity $Z = 0.0001$. Our binary star populations predict $\sim 37\%$ fewer thermally-pulsing AGB stars than our single star populations, leading to a $\sim 40\%$ decrease in the amount of ejected C and a $\sim 35-40\%$ reduction in elements synthesised through the slow neutron capture process. The uncertainty introduced by the mass-loss from stellar winds on the AGB makes the impact of binary evolution on the total amount of ejected N uncertain. The total N yield ejected by our binary star populations ranges from a $17\%$ to a $36\%$ decrease compared to our single star populations. However, our binary populations overproduce N by over an order of magnitude during the period $300-700\,$Myr after formation.
\end{abstract}

\section{INTRODUCTION }
\label{sec:intro}

Asymptotic giant branch (AGB) stars are evolved stars born with low to intermediate masses, $\sim 0.7-7\Msun$, depending on metallicity. AGB stars are essential for the chemical enrichment of the Universe, as they synthesise a significant portion of the C, N, F, and about half of the nuclides heavier than iron \citep{Kobayashi2020} through the slow-neutron capture process ($s$-process) \citep{Clayton1961, Lugaro2023}. In the early Universe, at metallicities of $Z \lesssim 0.0001$, AGB stars also contributed significantly to the Galaxy dust budget \citep{Valiante2009, Ventura2021, Yates2024} and to the production of Mg \citep{Fenner2003, Doherty2014_2}. 

The envelopes of AGB stars become enriched with heavy nuclides after the onset of repeated unstable shell He burning, known as thermal pulses. These thermal pulses drive structural change within the star, which allows for periodic episodes of stellar nucleosynthesis and convective mixing. Thermally-pulsing AGB (TP-AGB) stars synthesise nuclides such as C and F through partial shell He burning. These elements are convectively mixed into the outer stellar envelope during third dredge-up events, which can occur after a thermal pulse. Depending on the metallicity, TP-AGB stars with masses $\gtrsim 3\Msun$ may also experience temperatures at the bottom of their convective envelopes $\gtrsim 50\,$MK, which is sufficient for H-burning \citep{Boothroyd1995, Karakas2010}. H-burning at the bottom of the convective envelope is known as hot-bottom burning. Hot-bottom burning allows AGB stars to contribute significantly to the Galaxy's N budget. For detailed reviews on AGB evolution and nucleosynthesis, see \citet{Herwig2005} and \citet{Karakas2014}.

The primary site of s-process nucleosynthesis in AGB stars is the He-rich intershell between the H and He-burning shells. During a third dredge-up event, protons are transported into the He-rich intershell. These protons fuse with \iso{12}C, which then produces the neutrons needed for the s-process via the \iso{13}C($\alpha$,n)\iso{16}O reaction. In hot-bottom burning stars, H-burning during the third dredge-up prevents protons from mixing into the He-rich intershell \citep{Goriely2004}. The $s$-process can also be active in the He-rich intershell during thermal pulses when temperatures reach $> 300\,$MK, using neutrons synthesised via the \iso{22}Ne($\alpha,n$)\iso{25}Mg reaction \citep{Karakas2012, Lugaro2012}.

Mass-loss through stellar winds allows AGB stars to eject their nuclides into the interstellar medium. The total amount of an element or isotope ejected by a star or population over its lifetime is known as the stellar yield (see Section \ref{subsec:yields}). The stellar yields of AGB stars at $Z=0.0001$ (or $\text{[Fe/H]} \approx -2.2$ where $\text{[Fe/H]} \approx \text{log}_{\rm 10}[Z/Z_{\rm \odot}]$), are not well constrained. Accurate yields at such low metallicities are essential to interpret the chemical signatures of ancient stars and reconstruct the enrichment history of the early Milky Way and its satellite galaxies. Detailed stellar models that evolve stars by directly solving the equations of stellar evolution \citep{Herwig2004, Karakas2010, Cristallo2015, Ritter2018, Choplin2025} differ in their treatments of convective mixing and mass-loss, resulting in large variations in their stellar yields. Because the only surviving stars from the early Universe are born with masses $\lesssim 1\Msun$, it is challenging to constrain stellar models across a range of initial masses at this metallicity.

In Galactic chemical evolution, the chemical contributions of AGB stars are often calculated using stellar yields from single-star models \citep{Kobayashi2020, Prantzos2020}. However, observations of the remaining G, F, and K-type stars in the Galactic halo show that at least half of low- and intermediate-mass stars at $Z\sim0.0001$ exist in binaries \citep{Gao2014, Yuan2015}. Binary mechanisms such as Roche-lobe overflow \citep{Eggleton1983}, stellar wind accretion \citep{Bondi1944, Abate2013}, common envelope, and mergers, alter the evolutionary pathway of a star \citep{Iben1991, DeMarco2017}. This is evidenced by objects such as blue stragglers \citep{Bailyn1995, Leigh2013}, C-enhanced metal-poor stars \citep{Beers2005, Frebel2015, Sharma2018}, and He-core white dwarfs \citep{Cool1998, Serenelli2002}. 

The AGB is the final major nuclear-burning stage of low- and intermediate-mass stellar evolution, and is the phase in which most heavy elements are synthesised. At solar metallicity, it was found that the disruption of a stellar companion can reduce the ejected amount of C and $s$-process elements from a stellar population by up to $25\%$ \citep{Osborn2025}. Binary evolution can limit the ability of AGB stars to contribute to the chemical evolution of the Universe \citep{Izzard2004_Thesis}. Few Galactic chemical evolution models have used the stellar yields from low- and intermediate-mass synthetic binaries in their calculations \citep{DeDonder2002, DeDonder2004, Sansom2009, Yates2024}, however they discuss only a few key elements.

In this work, we use the binary population synthesis code \textsc{binary\_c} \citep{Izzard2004, Izzard2006, Izzard2009, Izzard2018, Izzard2023, Hendriks2023} to model and calculate the elemental yield of all stable elements up to Bi (excluding Li, B and Be) from low and intermediate-mass stellar populations at metallicity $Z=0.0001$ and quantify the impact introduced by binary evolution. Here we define low mass stars to have masses $\sim 0.7-3\Msun$ and intermediate mass stars to have masses $\sim 3-7\Msun$. We evolve five stellar model sets using various wind mass-loss prescriptions on the TP-AGB to reflect the varying treatments used in detailed AGB models \citep{Herwig2004, Karakas2010, Ritter2018}. We also calculate delay-time distributions of the ejected C, N, F, Sr, Ba, and Pb from our stellar populations.

This paper is structured as follows. Section \ref{sec:models} describes how we build our synthetic models and stellar populations with \textsc{binary\_c}, including updates to the treatment of the CO core mass (Section \ref{sec:CO_masses}) and the temperature at the base of the convective envelope (Section \ref{sec:Tbce}). In Section \ref{sec:Results}, we show the stellar yields from our stellar populations and describe the changes introduced by binary evolution. Section \ref{sec:Discussion} discusses our results and the uncertainty in the evolution of stars at $Z=0.0001$ and binary evolution. Finally, we highlight our conclusions in Section \ref{sec:Conclusion}.

\section{BINARY POPULATION SYNTHESIS MODELS}
\label{sec:models}

We use the binary population synthesis code \textsc{binary\_c} version 2.2.4, the latest official release at the time of writing, to model our stellar populations. We use \textsc{binary\_c} as it is the only binary population synthesis code that parameterises AGB stars with enough detail to also model AGB stellar nucleosynthesis (see \citealp{Izzard2006} for more details). This allows us to calculate the stellar yields directly from our modelled populations.

Our model parameters are presented in Table \ref{tab:parameters}. We choose an initial single and primary star mass range of $0.7-7\Msun$ as our single stars born with mass $\lesssim 0.7\Msun$ do not evolve off the main sequence during the $15\,$Gyr of simulation time, and stars of masses $\gtrsim 7 \Msun$ explode as supernovae and do not evolve through the TP-AGB. Initial chemical abundances lighter than \iso{76}Ge are estimated from \citet{Kobayashi2011} for $Z=0.0001$, and those including and heavier than \iso{76}Ge are scaled from the Solar abundances (where $Z = 0.0142$) presented in \citet{Asplund2009} to $Z=0.0001$. 

Results from the Galactic chemical evolution models from \citet{Kobayashi2011, Kobayashi2020} find that the stellar yields calculated from \citet{Karakas2010} match observations of N in the solar neighbourhood for ${\rm [Fe/H]} > -1.5$, where the contribution from AGB stars becomes dominant. Therefore, following the results from \citet{Karakas2010}, we set hot-bottom burning to occur in stars with masses $>3\Msun$. Additionally, the stars modelled in \citet{Karakas2010} were evolved until their envelope masses reduced to $\sim 0.1\Msun$, where they continued to experience efficient third dredge-up, allowing the continued enrichment of heavy nuclides in the stellar envelope. Therefore, we set our \textsc{binary\_c} models to terminate the third dredge-up at an envelope mass of $0.1\Msun$. 

To model $s$-process nucleosynthesis in \textsc{binary\_c}, we adopt the He-rich intershell abundance table described in \citet{Abate2015}, which is interpolated from the detailed models described in \citet{Lugaro2012} and includes 320 isotopes. During a third dredge-up event, the depth to which protons are transported into the He-rich intershell is uncertain. The detailed models from \citet{Lugaro2012} introduce a `partial mixing zone', defining the depth protons penetrate the He-rich intershell. In \citet{Abate2015_2}, they found that a partial mixing zone mass of $0.002\Msun$ at masses $\leq 3\Msun$ best reproduced the observed surface abundances of C-enhanced metal-poor stars. At masses $>3\Msun$, we set the mass of the partial mixing zone to be zero as H burning during the third dredge-up inhibits protons being transported into the He-rich intershell \citep{Goriely2004}.

\begin{table*} 
    \begin{threeparttable}
	\caption{Key stellar grid and model parameters shared by all model sets. Model parameters not listed here are set to the \textsc{binary\_c} V2.2.4 default. A complete list of model parameters may be obtained upon request from the corresponding author.}
	\centering
	\begin{tabular}{| c | c |}
    \hline
		\headrow Parameter & Setting \\
        \hline
		\midrule
		Initial single star mass, $M_{\rm 0}$, and primary star mass, $M_{\rm 1,0}$, range & $0.7-7 {\rm M_{\odot}}$ \\
		$M_{\rm 0}$ and $M_{\rm 1,0}$ grid-sampling probability distributions & Log-uniform in $M_{\rm 0}$ ($\times 1000$ sampled) and $M_{\rm 1,0}$ ($\times 100$ sampled)\\
        $M_{\rm 0}$ and $M_{\rm 1,0}$ birth probability distributions & \citet{Kroupa2001}, normalised between $0.01-150\Msun$ \\
		Initial secondary star mass, $M_{\rm 2,0}$ range & $0.1M_{\rm \odot} - M_{\rm 1,0}$ \\
        $M_{\rm 2,0}$ grid-sampling and birth probability distributions & Uniform in $M_{\rm 2,0}/M_{\rm 1,0}$ ($\times 100$ sampled)\\
        Initial orbital period, $p_{\rm 0}$, range & $1-10^6$ days \\
        $p_{\rm 0}$ grid-sampling and birth probability distributions & Log-uniform in $p_{\rm 0}$ ($\times 100$ sampled)\\
        Metallicity, $Z$ & $0.0001$ \\
        Simulation Time & $15 \,$Gyr \\
        Initial chemical abundance & \citet{Kobayashi2011} and \citet{Asplund2009} scaled to $Z=0.0001$ \\
        TP-AGB core, radius, and luminosity algorithms & \citet{Karakas2002} \\
        He-intershell abundance tables & \citet{Abate2015} \\
        Mass of the partial mixing zone in the He-rich intershell & 0.002 at $M \leq 3\Msun$ and no pmz at $M > 3\Msun$ \citep{Abate2015_2} \\
        Minimum mass for hot-bottom burning & $3\Msun$ \citep{Karakas2010} \\
        Minimum envelope mass for third dredge-up & $0.1\Msun$ \citep{Karakas2010} \\
        Common envelope energy binding parameter $\lambda_{\rm CE}$ & \citet{Dewi2000} \\
        Roche-lobe overflow treatment & \citet{Claeys2014} with thermal limit multiplier = 10 \\
        Wind Roche-lobe overflow treatment & \citet{Abate2013} q-dependent \\
        Wind angular momentum loss & Spherically symmetric \citep{Abate2013} \\
        Roche-lobe overflow angular momentum transfer model & Conservative \\
        Non-conservative angular momentum loss  & Isotropic \citep{Abate2013} \\
        
		\bottomrule
	\end{tabular}
	\label{tab:parameters}
    \end{threeparttable}
\end{table*}

To investigate the uncertainty introduced by stellar winds, we simulate stellar populations from five model sets evolved with various TP-AGB mass-loss prescriptions as described in Table \ref{tab:5pops}. For each model set, we produce a grid of $1000$ single and $10^6$ binary star models sampled as described in Table \ref{tab:parameters}. In model set $VW$, we apply the mass-loss prescription used in \citet{Karakas2002}, which is from \citet{Vassiliadis1993}. The \citet{Vassiliadis1993} mass-loss prescription is often used in other studies using \textsc{binary\_c}, including \citet{Abate2015}. Some detailed models use the mass-loss prescription from \citet{Bloecker1995} with $\eta$ values varying between 0.01 and 0.1, often estimated by extrapolating from higher metallicities \citep{Ventura2002, Herwig2004, Ritter2018}. Therefore, in model sets $B01$ and $B02$, we use mass-loss as described in \citet{Bloecker1995} with $\eta = 0.01$ and $0.02$, respectively. In \citet{Karakas2010}, they use mass-loss as described in \citet{Vassiliadis1993} for stars with masses $\leq 3\Msun$ and \citet{Reimers1975} for masses $> 3\Msun$ with $\eta$ values ranging from $5$ to $10$. Therefore, in model sets $VW\_B01$ and $VW\_B02$, we transition between the \citet{Vassiliadis1993} and \citet{Bloecker1995} mass-loss prescriptions. To facilitate the smooth transition between the TP-AGB mass-loss prescriptions at stellar mass around $3\Msun$, we use

\begin{equation}
    \label{eq:WindTransition}
    \dot M_{\rm TPAGB} = (1-f_{\rm l}) \dot M_{\rm VW93} + f_{\rm l} \dot M_{\rm B95} \text{,}
\end{equation}

\noindent where $\dot M_{\rm TPAGB}$ is the mass-loss during the TP-AGB, $\dot M_{\rm VW93}$ is the mass-loss calculated using the \citet{Vassiliadis1993} prescription, $\dot M_{\rm B95}$ is the mass-loss calculated using \citet{Bloecker1995}, and

\begin{equation}
    f_{\rm l} = \frac{1}{1 + 0.0001^{M_{\rm 1TP} - 3}},
\end{equation}

\noindent where $M_{\rm 1TP}$ is the total stellar mass in $\Msun$ at the first thermal pulse. We use the \citet{Bloecker1995} prescription for $M \gtrsim 3\Msun$, instead of the \citet{Reimers1975} prescription like in \citet{Karakas2010}, to avoid needing to also transition our mass-loss treatment near $3.5\Msun$ and $4.5\Msun$ to model how $\eta$ changes like in \citet{Karakas2010}.

\begin{table*} 
    \begin{threeparttable}
	\caption{AGB stellar wind prescriptions of our five model sets.}
	\centering
	\begin{tabular}{| c | c |}
        \hline
		\headrow Model Set & AGB Wind Prescription \\
		\midrule
		$VW$ & \citet{Vassiliadis1993} \\
        $B01$ & \citet{Bloecker1995} with $\eta = 0.01$  \\
        $B02$ & \citet{Bloecker1995} with $\eta = 0.02$ \\
        $VW\_B01$ & \citet{Vassiliadis1993} at $M \lesssim 3M$ and \citet{Bloecker1995} with $\eta = 0.01$ at $M \gtrsim 3M$ \\
        $VW\_B02$ & \citet{Vassiliadis1993} at $M \lesssim 3M$ and \citet{Bloecker1995} with $\eta = 0.02$ at $M \gtrsim 3M$ \\
		\bottomrule
	\end{tabular}
	\label{tab:5pops}
    \end{threeparttable}
\end{table*}

\subsection{CO Core Masses}
\label{sec:CO_masses}

In \textsc{binary\_c}, the CO core mass prior to the TP-AGB is calculated based on fits from \citet{Hurley2000} to the models described in \citet{Pols1998}. However, the CO core mass at the first thermal pulse is calculated using fits to \citet{Karakas2002}. A key difference between these models is that \citet{Pols1998} calculate their models with convective overshoot, whereas \citet{Karakas2002} do not. This results in the CO cores at the beginning of the EAGB from \citet{Pols1998} being up to about $0.4\Msun$ more massive for the same initial mass than those in \citet{Karakas2002}. This causes numeric issues in \textsc{binary\_c} when the CO core is more massive at the beginning of the EAGB than the predicted core mass at the first thermal pulse. At $Z=0.0001$, this occurs at masses between about $6-6.5\Msun$, and \textsc{binary\_c} responds by forcing the star to explode in a core-collapse supernova, despite the core lacking the mass to do so.

We employ a similar solution to that used in \citet{Osborn2023}. We refit the CO core masses at the beginning of the EAGB to those calculated in \citet{Karakas2010}, reducing the CO core mass at the beginning of the EAGB. Our resulting fit for the CO core mass, in $\Msun$, at the beginning of the EAGB, $M_{\rm CO, EAGB}$, is

\begin{equation}
\label{eq:EAGB_cores}
    \begin{split}
    &M_{\rm CO, EAGB} = \\
    &(1-f_2)\left[ (8.24 \times 10^{-3}) M_{\rm EAGB}^2 + (2.83\times10^{-2}) M_{\rm EAGB} + 0.244 \right] \\
    &+ f_2M_{\rm CO, Pols98},
    \end{split}
\end{equation}

\noindent where $M_{\rm EAGB}$ is the total mass at the beginning of the EAGB in $\Msun$ and $M_{\rm CO, Pols98}$ is the CO core mass as estimated using the fit to \citet{Pols1998} and 

\begin{equation}
    \label{eq:smoothEAGB}
    f_2 = \frac{1}{1 + 0.0001^{M_{\rm EAGB} - M_x}},
\end{equation}

where $M_x = 6\Msun$. Equation \ref{eq:smoothEAGB} smooths the transition between our fit to the CO core masses calculated in \citet{Karakas2010} and those calculated in \citet{Pols1998} at $M_x$. The \textsc{binary\_c} code uses a similar method to Equation \ref{eq:smoothEAGB} to transition between their fits to the CO core at the first thermal pulse, where they transition their fit to models described in \citet{Karakas2002} and \citet{Pols1998} at $M_x = 7\Msun$. For consistency, we set $M_x = 6\Msun$ for the transition of the treatment CO core mass at the first thermal pulse. The new fits eliminate the exploding EAGB stars, and results in $Z=0.0001$ stars with masses $\gtrsim6.2\Msun$ growing sufficiently massive cores to end their lives in a supernova. 

The reduced CO core masses at the beginning of the EAGB results in the radii and luminosities of our stars suddenly decreasing between the final time step of the core He burning phase and the first time step of the EAGB. However, there is no significant impact on the overall stellar evolution and yields calculated from our models. The luminosities and radii of our stars modelled with $M_{\rm CO, EAGB}$ fit to both \citet{Pols1998} and \citet{Karakas2010} finish the EAGB with near identical radii and luminosities. Note that \textsc{binary\_c} does not model any stellar nucleosynthesis using the CO core mass during the EAGB.

\subsection{Temperature at the base of the convective envelope}
\label{sec:Tbce}

The treatment of hot-bottom burning in \textsc{binary\_c} is detailed in \citet{Izzard2004, Izzard2006}. In \textsc{binary\_c}, the temperature at the base of the convective envelope, $T_{\rm bce}$, in Kelvin, is calculated using 

\begin{equation}
    \begin{split}
    {\rm log_{10}}(T_{\rm bce}) = f_{\rm Trise} \times {\rm log_{10}}(T_{\rm bce, max}) \times f_{\rm Tdrop},
    \end{split}
\end{equation}

\noindent where $T_{\rm bce,max}$ is the maximum $T_{\rm bce}$ calculated for the star (see Equation 37 of \citealp{Izzard2004}), $f_{\rm Trise}$ is the rise in temperature during the first few thermal pulses (see Equation 39 in \citealp{Izzard2004}), and 

\begin{equation}
    \label{eq:HBBfit_old}
    f_{\rm Tdrop} = (M_{\rm env}/M_{\rm env, 1TP})^{0.02},
\end{equation}

\noindent where $M_{\rm env}$ is the mass of stellar envelope and $M_{\rm env, 1TP}$ is the mass of the stellar envelope at the first thermal pulse (see Equation 40 from \citealp{Izzard2004}). 

\begin{figure}[hbt!]
\centering
\includegraphics[width=\linewidth]{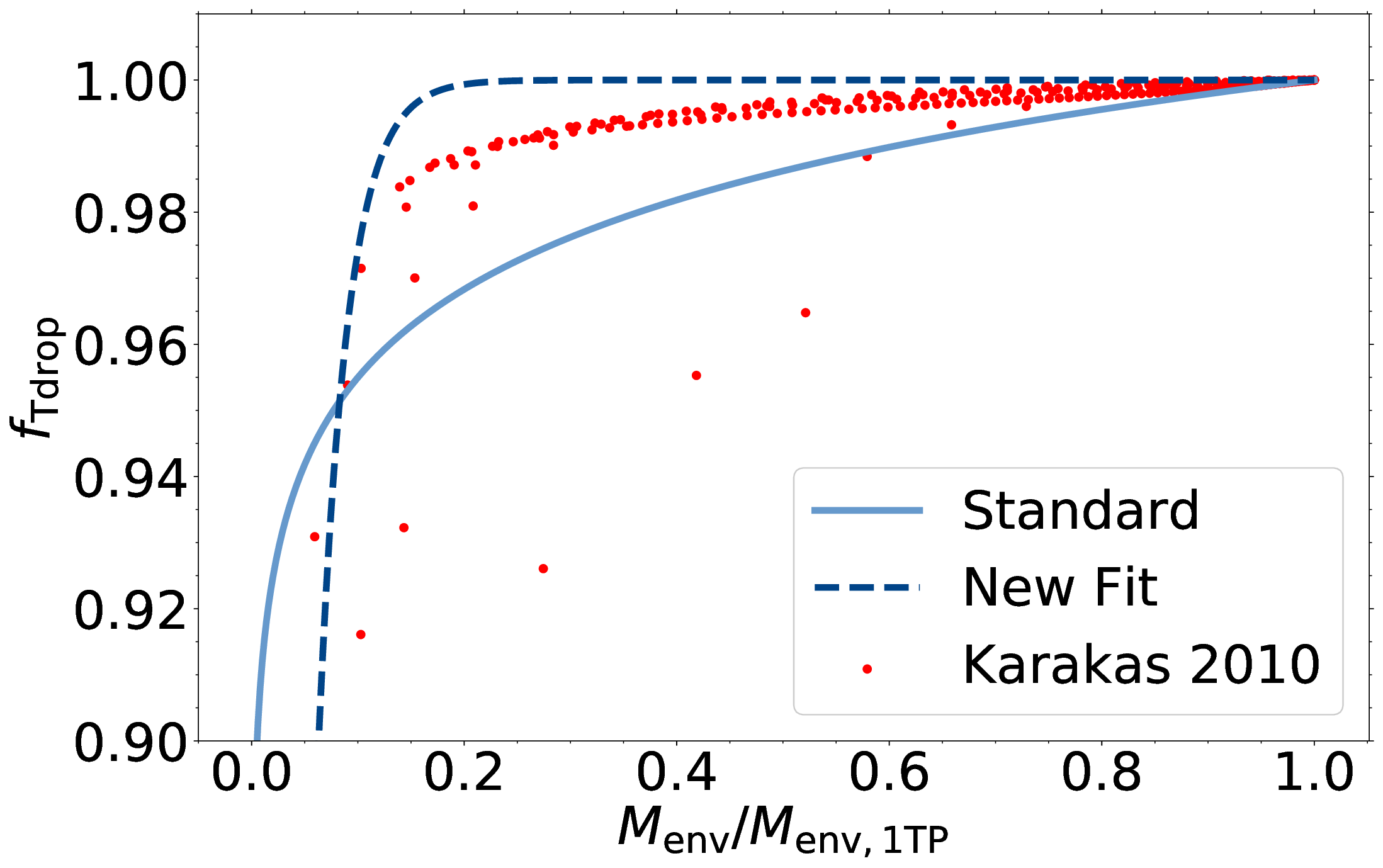}
\caption{We compare the fit for $f_{\rm Tdrop}$ described by Equation \ref{eq:HBBfit_old} (Standard) to our new fit described by Equation \ref{eq:HBBfit} (New Fit), and the models presented in \citet{Karakas2010}. We show $f_{\rm Tdrop}$ as a function of $M_{\rm env}/M_{\rm env, 1TP}$.}
\label{fig:HBB_cooling}
\end{figure}

Figure \ref{fig:HBB_cooling} shows the results of Equation \ref{eq:HBBfit_old} compared to results from the stellar-models of initial masses $\geq 3\Msun$ described in \citet{Karakas2010}, which predict sufficient temperatures for hot-bottom burning. In \textsc{binary\_c}, Equation \ref{eq:HBBfit_old} results in $T_{\rm bce}$ cooling too quickly as $M_{\rm env}/M_{\rm env,1TP}$ decreases compared to the stars modelled in \citet{Karakas2010}. This results in hot-bottom burning elements such as N being under-produced.

To improve the stellar yields of our stars modelled using \textsc{binary\_c} to better fit the results of \citet{Karakas2010}, we refit $f_{\rm Tdrop}$ to

\begin{equation}
    \label{eq:HBBfit}
    \begin{split}
    f_{\rm Tdrop} = 1-{\rm exp}\left(-\frac{M_{\rm env}/M_{\rm env, 1TP}}{0.027} \right).
    \end{split}
\end{equation}

Figure \ref{fig:HBB_cooling} shows that our new fit for $f_{\rm Tdrop}$ results in $T_{\rm bce}$ cooling more slowly with decreasing envelope compared to \citet{Karakas2010}. Equation \ref{eq:HBBfit} has a root mean squared error value of $5 \times 10^{-3}$ when considering $M_{\rm env}/M_{\rm env, 1TP} > 0.3$, which indicates a better fit to the data from \citet{Karakas2010} than Equation \ref{eq:HBBfit_old} which has a root mean squared error value of $7 \times 10^{-3}$. At $M_{\rm env}/M_{\rm env, 1TP} < 0.3\Msun$, Equation \ref{eq:HBBfit_old} is the better fit. However, this is not an issue since stars with $M_{\rm env}/M_{\rm env,1TP} < 0.3$ have less than $0.2\%$ of the TP-AGB phase remaining and $M_{\rm env}/M_{\rm env,1TP}$ is declining rapidly.

\subsection{Stellar and Population Yields}
\label{subsec:yields}
We calculate both stellar and population yields as described in \citet{Osborn2023, Osborn2025}, where we only consider the contribution of mass-loss due to stellar winds to the total yield. We calculate the total stellar yield using, 

\begin{equation}
    \label{eq:Yield}
    y_{i} =  \int ^{\tau} _0 X(i,t) \frac{{\rm d}M}{ {\rm d}t}{\rm d}t,
\end{equation}

\noindent where $y_{i}$ is the total stellar yield of element $i$ in $\Msun$, $X(i,t)$ is the surface mass fraction of species $i$, $\tau$ is the lifetime of the star, and $\frac{{\rm d}M}{ {\rm d}t}$ is the mass-loss rate from the stellar system noting it is always positive. We assume all short-lived radioactive isotopes have decayed. We do not decay the long-lived radioisotopes \iso{48}Ca, \iso{87}Rb, \iso{96}Zr, \iso{113}Cd, \iso{115}In, \iso{144}Nd, \iso{147}Sm, \iso{148}Sm, \iso{151}Eu, \iso{176}Lu, \iso{187}Re, \iso{186}Os, and \iso{209}Bi.

In our binary models, we calculate the stellar yield of the primary, secondary, and post-merger stars separately. Mass ejected via mass transfer, common envelopes, and mergers are also included in our stellar yield calculation and are treated as in \citet{Osborn2023}. We assume that material ejected during a mass transfer or common envelope event originates from the donor star. During a stellar merger event, we assume the ejected material is a mix of both stellar envelopes, depending on the evolutionary phases of the stellar components. For example, if a giant star merges with a main-sequence star following a common envelope event, we assume the ejected material originates from the donating giant star (see \citealp{Osborn2023} for more details). We calculate the stellar yield contribution from stellar winds, mass-transfer, and stellar mergers over a total of $15\,$Gyr simulation time.

For Galactic chemical evolution, it is important to determine the net production or destruction of any given element. The net yield, $y_{i, {\rm net}}$, of a given species, $i$, is defined as,

\begin{equation}
    \label{eq:NetYield} 
    y_{i, {\rm net}} =  \int ^{\tau} _0 \left[ X(i,t) - X(i,0) \right] \frac{{\rm d}M}{ {\rm d}t}{\rm d}t,
\end{equation}

\noindent where $X(i,0)$ is the surface mass fraction of species $i$ at birth.

 To express the total or net yield contribution of each model to a stellar population in units $\Msun$ per $\Msun$ of star-forming material ($\Msun/{\rm M_{\odot,SFM}}$), we apply a weighting factor $w_j$ to each model $j$ in our model sets where

\begin{equation}
    \label{eq:Weighting_S}
    w_{j,{\rm s}} = (1-f_{\rm b}) \frac{w_{\rm m}}{n_{\rm s}} \frac{\pi_{\rm s}(\mathbf{x}_{j})}{\xi_{\rm s}(\mathbf{x}_{j})},
\end{equation}

\noindent for the single-star portion of the population and 

\begin{equation}
    \label{eq:Weighting_B}
    w_{j,{\rm b}} = f_{\rm b} \frac{w_{\rm m}}{n_{\rm b}} \frac{\pi_{\rm b}(\mathbf{x}_{j})}{\xi_{\rm b}(\mathbf{x}_{j})},
\end{equation}

\noindent for the binary star portion of the population where $f_{\rm b}$ is the binary fraction of the stellar population, $n_{\rm s}$ and $n_{\rm b}$ are the number of models sampled for our single and binary grids respectively, $\pi_{\rm s}(\mathbf{x}_{j})$ and $\pi_{\rm b}(\mathbf{x}_{j})$ respectively describe the theoretical probability distributions of initial conditions of the observed single and binary populations, and $\xi_{\rm s}(\mathbf{x}_{j})$ and $\xi_{\rm b}(\mathbf{x}_{j})$ are the probability distributions of our single and binary models, respectively, sampled in \textsc{binary\_c} \citep{Broekgaarden2019, Kemp2021, Osborn2025}, and $w_{\rm m}$ is a mass normalisation term describing the average number of stellar systems forming per $\Msun$ of star-forming material where,

\begin{equation}
    \label{eq:mass_norm_intro}
    w_{\rm m} = \frac{\int^{M_{\rm 1,max}}_{M_{\rm 1,min}} \pi(M_{\rm 1,0})\,{\rm d}M_{\rm 1,0}}{\int^{M_{\rm 1,max}}_{M_{\rm 1,min}} M_{\rm 1,0}\, \pi(M_{\rm 1,0})\,{\rm d}M_{\rm 1,0} + f_{\rm b}\int^{M_{\rm 2,max}}_{M_{\rm 2,min}} M_{\rm 2,0}\, \pi(M_{\rm 2,0})\,{\rm d}M_{\rm 2,0}},
\end{equation}

where $M_{\rm 1,0}$ is the initial mass of our single and binary primary stars born with a mass distribution $\pi(M_{\rm 1,0})$ normalised between $M_{\rm 1,min}$ and $M_{\rm 1,max}$ as described in Table \ref{tab:parameters}, $M_{\rm 2,0}$ is the initial mass of our secondary stars born with a mass distribution $\pi_{\rm b}(M_{\rm 2,0})$ normalised between $M_{\rm 2,min}$ and $M_{\rm 2,max}$ as described in Table \ref{tab:parameters}. The term $\int^{M_{\rm 1,max}}_{M_{\rm 1,min}} \pi(M_{\rm 1,0})\,{\rm d}M_{\rm 1,0}$ describes the total number of stellar systems forming in our population, $\int^{M_{\rm 1,max}}_{M_{\rm 1,min}} M_{\rm 1,0}\, \pi(M_{\rm 1,0})\,{\rm d}M_{\rm 1,0}$ is the total mass of the combined single and binary primary stars in our population, and $f_{\rm b}\int^{M_{\rm 2,max}}_{M_{\rm 2,min}} M_{\rm 2,0}\, \pi(M_{\rm 2,0})\,{\rm d}M_{\rm 2,0}$ describes the contribution of the binary secondary stars to the total mass of our stellar population.

The birth distributions of stars within the Galactic halo are uncertain \citep{Hallakoun2021, vanOirschot2014}. For our populations, we use the birth distributions for initial single star and primary mass, initial secondary mass, and initial orbital period as described in \citet{Osborn2025}, and summarised in Table \ref{tab:parameters}. 

We calculate the weighted total or net stellar yield $y_{\rm pop, i}$ of a given species $i$, in units $\Msun/{\rm M_{\odot,SFM}}$ of our mixed stellar population using

\begin{equation}
    \label{Eq:WYield}
    y_{\rm pop, i} = \sum^{n_{\rm s}}_{j=0} w_{j, {\rm s}} \times y_{i,j,{\rm s}} + \sum^{n_{\rm b}}_{j=0} w_{j, {\rm b}} \times (y_{i,j,{\rm b1}} + y_{i,j,{\rm b2}} + y_{i,j,{\rm b3}}),
\end{equation}

\noindent where $y_{i,j,{\rm s}}$ is the total or net stellar yield of element $i$ from each single star model $j$ in our model set and ${y_{i,j,{\rm b1}}}$, ${y_{i,j,{\rm b2}}}$, and ${y_{i,j,{\rm b3}}}$ are the total or net yields from our binary primary, secondary, and post-merger stars, respectively. In this work, we calculate the weighted total stellar yields of our populations with binary fractions ranging from 0 to 1 in increments of 0.1. The impact of stellar explosions, such as novae and supernovae, on the stellar and population yields is beyond the scope of this project. For stellar systems where an explosion occurs, we only use the contribution from stellar winds to the total yields.

\section{RESULTS}
\label{sec:Results}

In this section, we first compare our single stars' C and N total yields to detailed models. We then compare the weighted total yield of all stable elements between our single and binary star populations. Finally, we present delay-time distributions of the net C and N ejected by our single and binary populations.

\subsection{Single Star Yields}
\begin{figure*}[hbt!]
\centering
\includegraphics[width=\linewidth]{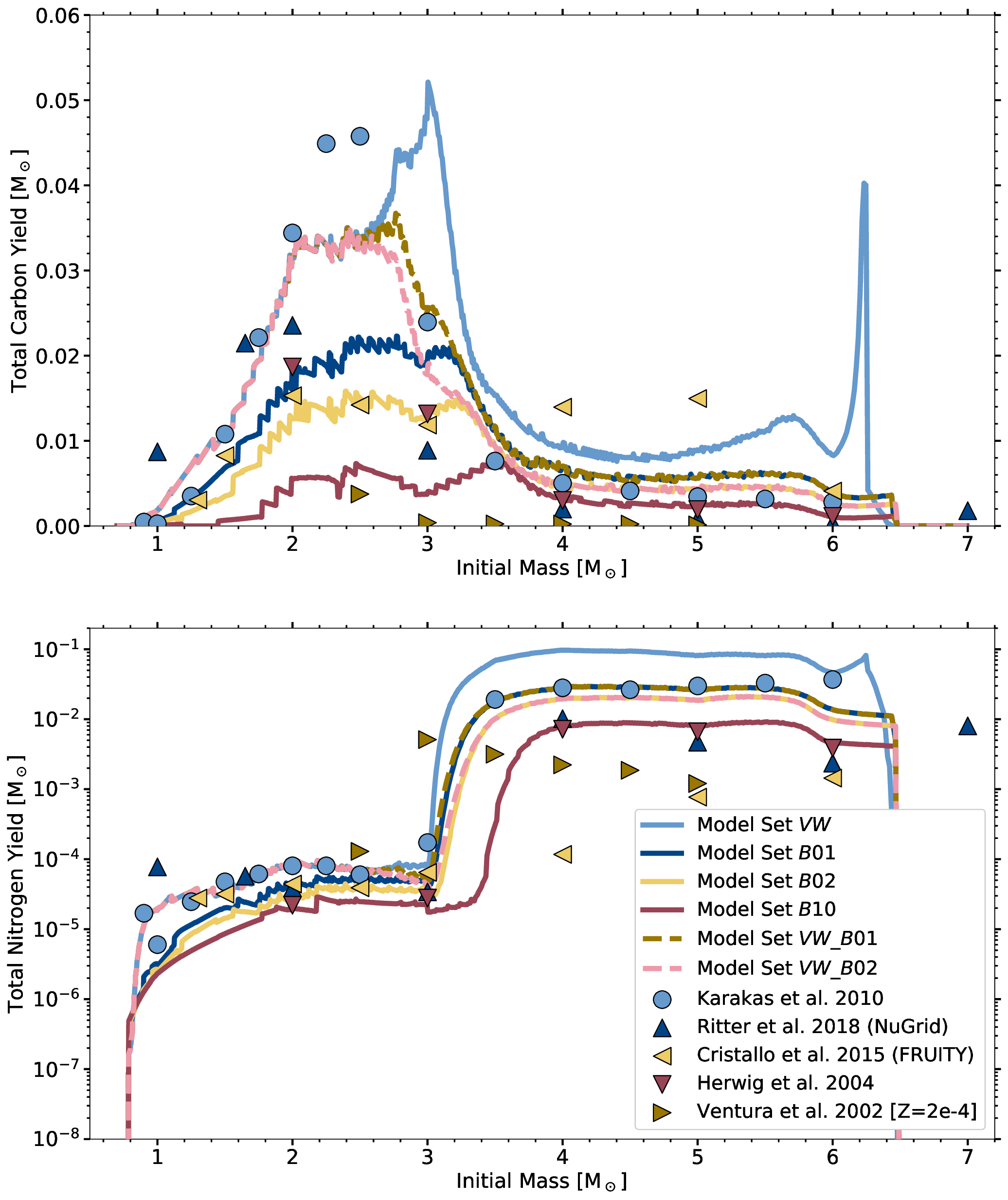}
\caption{Total stellar yield of C (top) and N (bottom) as a function of initial stellar mass. Here, we compare the results from detailed stellar evolution codes to those from our single stars models from our model sets as described in Table \ref{tab:5pops}. All results from detailed stellar evolution codes are calculated with $Z=0.0001$, except for \citet{Ventura2002} which uses $Z=0.0002$. Model set $B10$ describes our models where mass-loss on the TP-AGB is calculated using \citet{Bloecker1995} with $\eta = 0.1$.}
\label{fig:SingleCompare}
\end{figure*}

Figure \ref{fig:SingleCompare} shows the total C and N yields of our single star models compared to the yields calculated in \citet{Karakas2010}, \citet{Ritter2018}, \citet{Cristallo2015}, \citet{Herwig2004}, who all use $Z=0.0001$, and \citet{Ventura2002}, who uses $Z = 0.0002$. We also include the total C and N yields ejected by our single star models where mass-loss is modelled using \citet{Bloecker1995} with $\eta = 0.1$ on the TP-AGB, which is notated as model set $B10$.

Despite differing treatments of mass-loss on the TP-AGB, Populations $VW93\_B01$ and $VW93\_B02$ agree reasonably well with the C and N yields from \citet{Karakas2010}, as expected with our model calibrations. Our models disagree most with \citet{Ventura2002} and \citet{Cristallo2015}. The models from \citet{Cristallo2015} experience hot-bottom burning at masses $\geq 5\Msun$, which is more massive than our stars modelled in \textsc{binary\_c}, reducing the total N output of their stars. C yields from the models described in \citet{Ventura2002} are distinctly lower compared to the other models shown here. This is attributed to their relatively low third dredge-up efficiency of $0.3-0.5$, compared to the $\sim0.9$ in our models. 

From model set $B10$, the high mass-loss rates introduced using $\eta = 0.1$ result in the stellar envelopes of all single stars $<1.8\Msun$ being ejected before experiencing five thermal pulses. For all other model sets, stars $>0.9\Msun$ experience at least five thermal pulses. Models from \citet{Herwig2004}, who use \citet{Bloecker1995} with $\eta = 0.1$, do not model stars $<2\Msun$, and the C yields calculated for the $2\Msun$ and $3\Msun$ stars from \citet{Herwig2004} better agree with our model sets $B01$ and $B02$. Although set $B10$ reasonably reproduces the C and N yields from \citet{Herwig2004} for stars with masses $\gtrsim 3.5\Msun$, stars of mass $<3.5\Msun$ make up the majority of a stellar population \citep{Salpeter1955, Kroupa2001}. We therefore exclude the model set $B10$ from further analysis.

\subsection{Population Yields including Binaries}

Before we discuss the population yields, it is important to understand how binary evolution changes the evolution of individual stars. Table \ref{tab:SB_Compare} shows the formation rates of TP-AGB stars, including hot-bottom burning stars, and the total amount of material ejected by our single and binary populations from all model sets per unit of $M_{\rm \odot, SFM}$ using Equations \ref{eq:Weighting_S} and \ref{eq:Weighting_B}. We identify TP-AGB stars that experience at least five thermal pulses, and we identify hot-bottom burning stars with a total mass of at least $3.25\Msun$ at the fifth thermal pulse. Our stellar models are set up to have hot-bottom burning at masses $> 3\Msun$, but we make a conservative estimate to account for binary evolution. For our binary-star population, we include the contribution from the binary primary, secondary, and post-merger stars in our calculations. We highlight that the results of our binary population include the combined effects of binary evolution and the redistribution of the star-forming mass of our single-star populations into our secondary stars. Due to the formation of the secondary stars, the stellar mass distribution of our binary-star populations are bottom-heavy compared to our single-star populations. To examine the impact of redistributing star-forming mass into our secondary stars, independent of binary evolution, Table \ref{tab:SB_Compare} also includes results for our binary populations where the binary primary and secondary stars are treated as if they are single.

\begin{table} 
    \begin{threeparttable}
	\caption{Here we show the average of the total mass of material ejected by the single and binary populations calculated from our five model sets. We also show the average number of TP-AGB and hot-bottom burning TP-AGB stars forming in these populations. The population notated as `Binary$^{*}$' shows the results of our binary populations where we treat the binary primary and secondary stars as single stars.} The uncertainty is one standard deviation of the average.
	\centering
	\begin{tabular}{| c | c | c | c |}
        \hline
		\headrow Population & \makecell{Ejected material \\ $\Msun/{\rm M_{ \odot, SFM}}$} & \makecell{TP-AGB stars \\ per ${\rm M_{ \odot, SFM}}$} & \makecell{Hot-bottom burning \\ TP-AGB stars \\ per ${\rm M_{ \odot, SFM}}$} \\
		\midrule
		Single & $0.205 \pm 0.007$ & $0.20 \pm 0.01$ & $(1.96 \pm 0.05) \times 10^{-2}$ \\
        Binary & $0.215 \pm 0.005$ & $0.127 \pm 0.008$ & $(1.34 \pm 0.03) \times 10^{-2}$ \\
        Binary$^*$ & ${0.194 \pm 0.006}$ & ${0.19 \pm 0.01}$ & ${(1.73 \pm 0.05) \times 10^{-2}}$ \\
		\bottomrule
	\end{tabular}
	\label{tab:SB_Compare}
    \end{threeparttable}
\end{table}

We find our binary population produces about $37\%$ fewer TP-AGB stars per ${\rm M_{ \odot, SFM}}$ over the ${15\,}$Gyr simulation time, including $\sim 32\%$ fewer TP-AGB stars with hot-bottom burning than our single star population. Therefore, fewer stars are available to contribute C, N, and s-process elements to the interstellar medium. Our binary population also ejects about $\sim5\%$ more material per ${\rm M_{\odot, SFM}}$ than our single-star populations.

Table \ref{tab:SB_Compare} shows that the impact of redistributing star-forming mass into our binary secondary stars has minimal impact on the total number of TP-AGB stars in our binary population, as the average TP-AGB formation rate agrees with our single-star populations within one standard deviation. Therefore, we attribute the $37\%$ decrease in the formation of TP-AGB stars in our binary populations from our single-star populations to binary evolution. Table \ref{tab:SB_Compare} also shows that the formation of hot-bottom burning TP-AGB stars is more sensitive to the redistribution of star-forming mass into our secondary stars, with $12\%$ fewer hot-bottom burning TP-AGB stars forming than our single-star populations. This accounts for about $37\%$ of the missing stars from our binary-star populations with binary evolution.

\begin{figure}[hbt!]
\centering
\includegraphics[width=\linewidth]{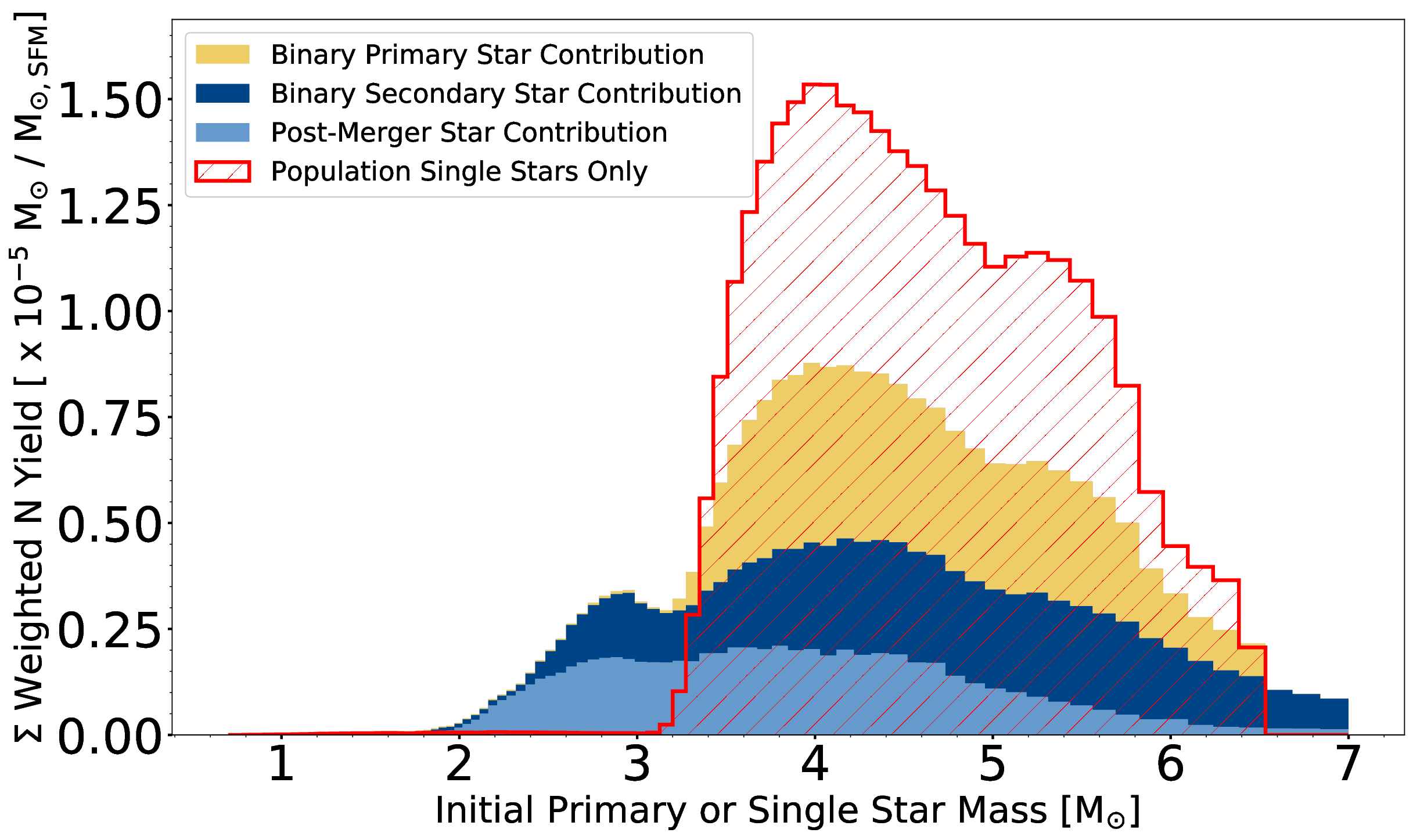}
\caption{Here we compare the population N yields of our single and binary star populations, calculated from model set $B02$. For our binary population, we show the contribution of the binary primary, secondary, and post-merger stars to the total population N yield. We show these results as a function of the initial single or binary-star mass. We bin the yield contribution of our secondary and post-merger stars by the initial mass of their binary primary stars. We stack the contributions from each component of the binary population, with their summation equalling the total population yield.}
\label{fig:N14Change}
\end{figure}

We now examine how including binaries in our population influences the yields. For example, in Figure \ref{fig:N14Change} we compare the N yield from our single-star (binary fraction of 0) and binary-star (binary fraction of 1) populations. These yields are calculated from the model set $B02$ using Equation \ref{Eq:WYield}. The total population yield from our binary population is 25\% lower compared to our single-star population. We can see from Figure \ref{fig:N14Change} that there is an overall reduction in the N ejected by stellar systems with primary or single star mass $\gtrsim 3\Msun$, reflecting the reduction in the formation of hot-bottom burning stars due to binary evolution. However, binary systems with primary masses $\lesssim 3\Msun$ overproduce N compared to our single star population. The additional N originates from our secondary stars, which accrete material through either mass-transfer or wind Roche-lobe overflow, and our post-merger objects. These stars enter the TP-AGB with masses $\gtrsim 3\Msun$, which allows the bottom of their convective envelopes to reach temperatures sufficient for hot-bottom burning. 

Table \ref{tab:WYields} shows weighted population yields (see Equation \ref{Eq:WYield}) for C, N, and Pb from all of our stellar populations. At a binary fraction of 1.0, we find a $35-40\%$ decrease in the ejected C, a $17-36\%$ decrease in the ejected N, and a $36-41\%$ decrease in the ejected Pb from all populations compared to our single star populations. 

\begin{table*} 
    \begin{threeparttable}
	\caption{Total population yields for all elements at binary fractions ranging from 0 to 1 for all model sets. Here, we show our results for C, N, and Pb. Tables showing the net and total stellar yields of all stable elements up to and including Bi, excluding Li, B, and Be, are available online.}
	\centering
	\begin{tabular}{| c | c | c | c | c | c | c | c | c | c | c | c | c |}
        \hline
		\headrow Element & Model Set & \multicolumn{11}{c}{Binary Fraction of Population} \\
        \headrow & & 0.0 & 0.1 & 0.2 & 0.3 & 0.4 & 0.5 & 0.6 & 0.7 & 0.8 & 0.9 & 1.0 \\
        \hline
		\midrule
	    C ($\times 10^{-3} \Msun/{\rm M_{\odot,SFM}}$) & $VW$ & 2.3 & 2.2 & 2.1 & 2.0 & 1.9 & 1.8 & 1.7 & 1.6 & 1.5 & 1.5 & 1.4 \\
        & $B01$ & 1.3 & 1.2 & 1.2 & 1.1 & 1.1 & 1.0 & 0.95 & 0.91 & 0.87 & 0.84 & 0.80 \\
        & $B02$ & 0.88 & 0.83 & 0.80 & 0.76 & 0.72 & 0.69 & 0.66 & 0.64 & 0.61 & 0.59 & 0.56 \\
        & $VW\_B01$ & 2.1 & 2.0 & 1.9 & 1.8 & 1.7 & 1.6 & 1.5 & 1.5 & 1.4 & 1.3 & 1.3 \\
        & $VW\_B02$ & 2.0 & 1.9 & 1.8 & 1.7 & 1.6 & 1.5 & 1.5 & 1.4 & 1.3 & 1.3 & 1.2 \\
        \hline 
        N ($\times 10^{-4} \Msun/{\rm M_{\odot,SFM}}$) & $VW$ & 16 & 15 & 15 & 14 & 13 & 13 & 12 & 12 & 11 & 11 & 10 \\
        & $B01$ & 4.7 & 4.5 & 4.4 & 4.3 & 4.1 & 4.0 & 3.9 & 3.8 & 3.7 & 3.6 & 3.5 \\
        & $B02$ & 3.2 & 3.1 & 3.0 & 2.9 & 2.8 & 2.7 & 2.6 & 2.6 & 2.5 & 2.5 & 2.4 \\ 
        & $VW\_B01$ & 4.8 & 4.6 & 4.5 & 4.4 & 4.3 & 4.2 & 4.1 & 4.0 & 3.9 & 3.8 & 3.8 \\
        & $VW\_B02$ & 3.2 & 3.2 & 3.1 & 3.0 & 3.0 & 2.9 & 2.8 & 2.8 & 2.8 & 2.7 & 2.7 \\
        \hline
        Pb ($\times 10^{-8} \Msun/{\rm M_{\odot,SFM}}$) & $VW$ & 1.7 & 1.6 & 1.6 & 1.5 & 1.4 & 1.3 & 1.3 & 1.2 & 1.1 & 1.1 & 1.0 \\
        & $B01$ & 1.1 & 1.0 & 0.97 & 0.91 & 0.87 & 0.83 & 0.79 & 0.75 & 0.71 & 0.68 & 0.65 \\
        & $B02$ & 0.71 & 0.68 & 0.65 & 0.61 & 0.59 & 0.56 & 0.53 & 0.51 & 0.49 & 0.47 & 0.45 \\
        & $VW\_B01$ & 1.7 & 1.6 & 1.5 & 1.4 & 1.4 & 1.3 & 1.2 & 1.2 & 1.1 & 1.1 & 1.0 \\ 
        & $VW\_B02$ & 1.7 & 1.6 & 1.5 & 1.4 & 1.4 & 1.3 & 1.2 & 1.2 & 1.1 & 1.1 & 1.0\\
        \hline
		\bottomrule
	\end{tabular}
	\label{tab:WYields}
    \end{threeparttable}
\end{table*}

\begin{figure*}[hbt!]
\centering
\includegraphics[width=\linewidth]{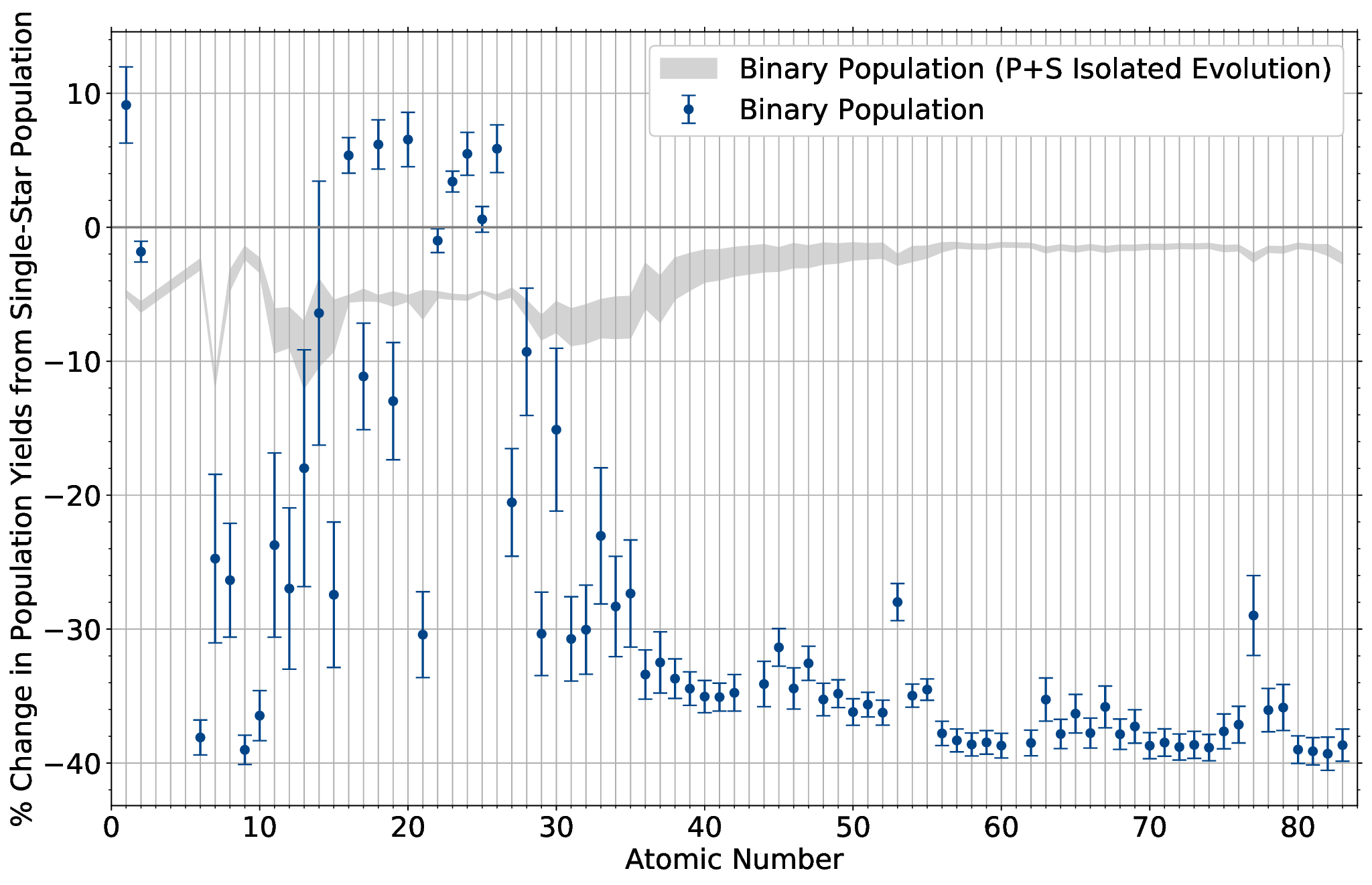}
\caption{Here we show the average of the percentage change in the total elemental yields of our binary star populations from our single star populations from our five model sets. For the data labelled `Binary Population', we are comparing our populations with a binary fraction of 1 to populations with a binary fraction of 0. The error bars indicate one standard deviation of the average, highlighting the variation introduced by our choice of mass-loss on the TP-AGB. For the data labelled `Binary Population (P+S Isolated Evolution)', we are showing the average and one standard deviation of our results where we evolve the stellar components of our binary-star population as if they are single.}
\label{fig:avgChange}
\end{figure*}

Here, we examine how the inclusion of binary stars influences the population yields of all studied elements. Figure \ref{fig:avgChange} shows the average percentage deviation in the binary population yields from the single star population yields for our model sets (see Table \ref{tab:5pops}). We show all elements with atomic numbers up to and including Bi, excluding Li, B, Be, and radioactive Tc and Pm. The change in Li ejected by our binary population compared to our single-star population varies from a $29\%$ decrease (Population $B01$) to a $230\%$ increase (Population $VW93$). Li yields calculated from stellar models are notoriously sensitive to the treatment of convective mixing and mass-loss \citep{Ventura2010, Lau2012, Gao2022}, and modelling the Cameron-Fowler mechanism \citep{Cameron1977} requires a level of detail not captured by our synthetic models, so we conclude that our Li results are unreliable. We exclude B and Be as they are not included in our nuclear network. Tc and Pm have no stable isotopes, and we add their contributions to the yields of their daughter nuclei. As with the results of Table \ref{tab:SB_Compare}, we also show our results of our binary populations where we evolve the primary and secondary stars as single stars, effectively turning off binary evolution, to indicate the dependence of redistributing the star-forming material of our population into the secondary stars on our results.

Figure \ref{fig:avgChange} shows that binary evolution has a high impact on the production of C, F, and Ne, with our binary-star populations producing $\lesssim40\%$ less than our single-star populations. C, F, and \iso{22}Ne are synthesised in the He-burning shells of TP-AGB stars, and the third dredge-up mixes these products into the stellar envelope. The reduction of C, F, and Ne is mainly attributed to binary evolution preventing the formation of TP-AGB stars. The decrease in TP-AGB systems also reduces the chemical yield of the s-process elements by about $35-40\%$. The weighted yields of the iron peak elements slightly increase in our binary population, but this is due to the increase in ejected material per M$_{\rm \odot,SFM}$ from our binary stars, as shown in Table \ref{tab:SB_Compare}, rather than any increase in elemental production. 

Elements synthesised through hot-bottom burning, such as N and Na, are underproduced by our binary populations compared to our single-star populations. Our choice of mass-loss prescription drastically alters the lifetimes of intermediate-mass TP-AGB stars, hence the large uncertainty on the yields of hot-bottom burning elements. For example, a single $5\Msun$ star modelled using mass-loss on the TP-AGB described with the \citet{Vassiliadis1993} prescription exists on the TP-AGB for $1.0 \times 10^6$ years, but the star modelled using the \citet{Bloecker1995} prescription with $\eta = 0.02$ exists for only $3.4 \times 10^5$ years, which is a $66\%$ decrease. Additionally, the effect of allocating star-forming material to our secondary stars introduces a comparable decrease in the production of hot-bottom burning elements as binary evolution. This is especially apparent for the yields of Al, Si and Ni where the average yields between our binary populations with and without binary evolution agree within one standard deviation. Our single-star models show that the production of Al, Si, and Ni, peaks in stars of masses $4-6\Msun$, which is the mass range most heavily impacted by our redistribution of star-forming mass into our secondary stars. However, low- and intermediate-mass stars do not contribute significantly to the Al, Si, and Ni in the Galaxy \citep{Kobayashi2020}.

The uncertainty in the lifetime of the TP-AGB also introduces uncertainty in the yields of the elements of the first s-process peak, such as Sr and Y. Models predict that in hot-bottom burning stars, the $s$-process is active during thermal pulses using neutrons produced via the \iso{22}Ne($\alpha,n$)\iso{25}Mg reaction. A single $5\Msun$ star modelled using mass-loss on the TP-AGB as described in \citet{Vassiliadis1993} experiences 158 third dredge-up events, but only experiences 42 when modelled using the \citet{Bloecker1995} prescription with $\eta = 0.02$.

\subsection{Delay-Time Distributions}
\label{sec:DTD}

\begin{figure}[hbt!]
\centering
\includegraphics[width=\linewidth]{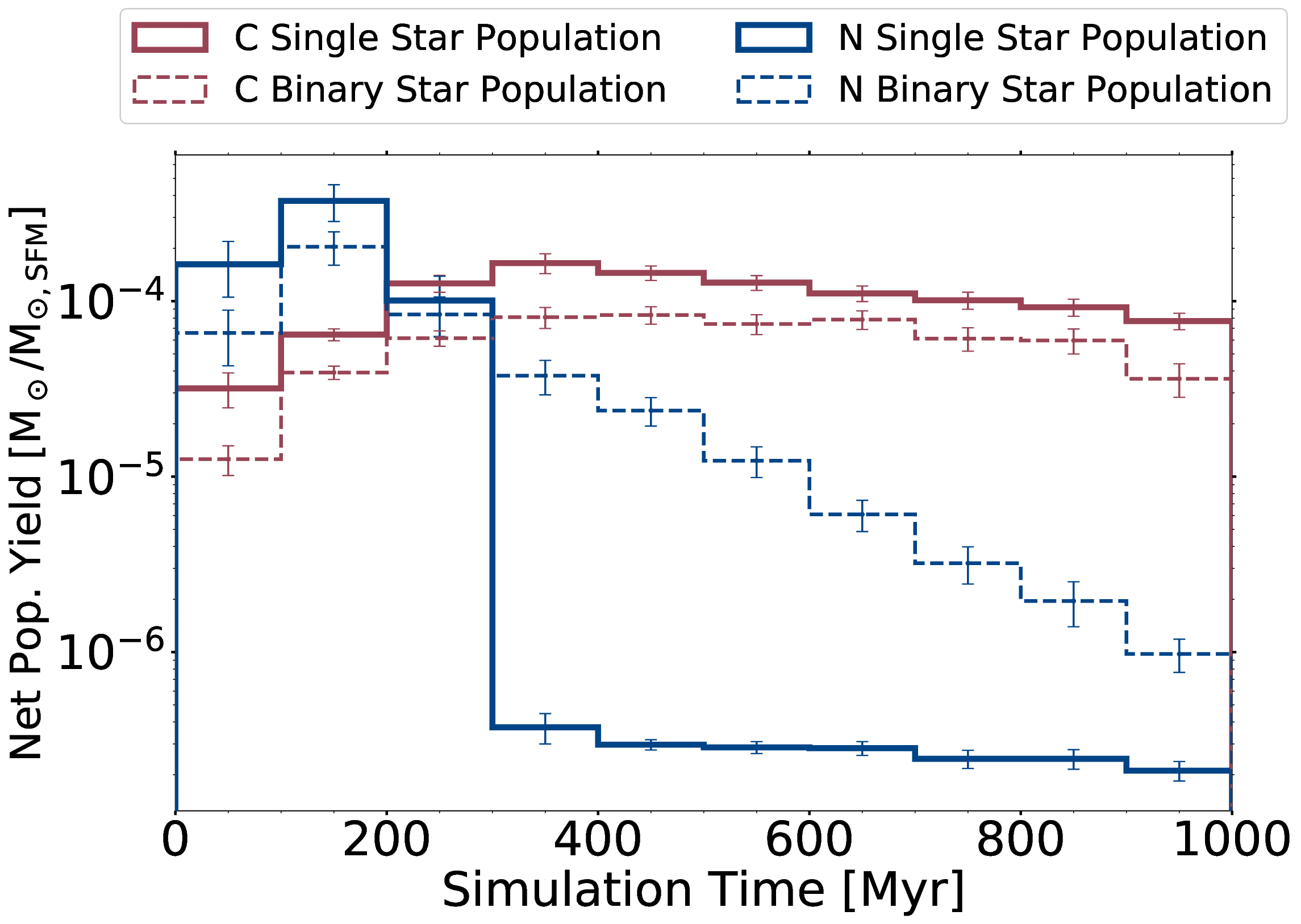}
\caption{Here, we show the average net C and N yield of our stellar populations as a function of time. We are comparing our populations where the binary fraction is 0 (single star population) and 1 (binary star population) following a single burst of star formation. We show our results up to $1\,$Gyr after formation, and we bin with a $100\,$Myr time-step. The histograms are transparent and overlapping. The error bars indicate one standard deviation in the average population yield, calculated from our five model sets.}
\label{fig:CN_Delay}
\end{figure}

It is important to the field of Galactic chemical evolution that we investigate how binaries influence the elemental production as a function of time. Figure \ref{fig:CN_Delay} shows the average net C and N ejected by our populations at binary fractions 0 and 1, in the first 1 Gyr following a burst of star formation. Note that these results reflect the combined effect of binary evolution and the redistribution of star-forming material of our population into the secondary stars. Tables showing the net C, N, F, Sr, Ba, and Pb ejected during the first $5\,$Gyr after formation for populations of binary fractions varying from 0 to 1 for all model sets are available online. 

Throughout the first Gyr, the introduction of binaries results in a consistent underproduction of C. For N, there is an underproduction in the first $\sim 300$ Myr as our binary populations produce fewer hot-bottom burning stars through binary evolution and the formation of secondary stars (see Table \ref{tab:SB_Compare}). However, between $300-700\,$Myr, our binary populations overproduce N by over an order of magnitude. After $700\,$Myr our binary populations continue to overproduce N by a factor of at least 2. The overproduction is mainly attributed to binary systems with initial primary mass $< 3\Msun$ (see Figure \ref{fig:N14Change}). Stellar mergers and mass transfer between the stars in these systems allow their stars to gain sufficient mass for hot-bottom burning. 

Figure \ref{fig:CN_Delay} shows the uncertainty introduced by our choice of mass-loss on the TP-AGB. At simulation times between $100-600\,$Myr, the underproduction of C introduced by binaries is significant to at least two standard deviations. In the case of N, our binary populations overproduce N compared to our single star populations at simulation times $\gtrsim 300\,$Myr after formation, significant to at least two standard deviations. These results indicate that binary evolution can significantly impact the yield outputs of C and N as a function of time.

\section{DISCUSSION}
\label{sec:Discussion}

Here, we compare our results to previous studies of populations of AGB stars in binaries. We then discuss the uncertainty in our intermediate-mass and binary models, our choice to exclude novae and supernovae from our population yields, and the limitations of comparing our results to observations.

\subsection{Comparison with Previous Work}

Galactic chemical evolution models that explore the impact of binary evolution include \citet{DeDonder2002, DeDonder2004} and \citet{Yates2024}. Here, we discuss their conclusions regarding their stellar wind contribution from low and intermediate-mass stars in comparison to what we find from our results.

We first start with a comparison to \citet{DeDonder2002} as they include the contribution from low- and intermediate-mass stars, noting they only do so for models of $Z \ge 0.001$. \citet{DeDonder2002} find that the inclusion of intermediate-mass binary stars results in a reduction in the C yielded from their populations compared to their single-star populations, and binarity has a negligible impact on the yields of their low-mass stars. This disagrees with our populations, as low-mass stars are the primary source of C in our populations. The low-mass stars modelled in \citet{DeDonder2002} are reported to only contribute He to the interstellar medium, and therefore likely do not experience any third dredge-up. Additionally, the models used in \citet{DeDonder2002} do not model hot-bottom burning, and their models do not reproduce the N abundances observed in the Solar neighbourhood for $\text{[Fe/H]} < -1$. The study presented in \citet{DeDonder2004} update their low- and intermediate-mass stellar yields based on the models calculated in \citet{vandenHoek1997} to include hot-bottom burning stars, and they still find that binary evolution reduces the C contribution from their low- and intermediate-mass populations. Although the C now originates from low-mass stars rather than intermediate-mass stars as in \citet{DeDonder2002}.

The work from \citet{Yates2024} build their stellar populations using models from \textsc{binary\_c}. They define a `wind group' which describes the combined contribution by stellar winds, Roche-lobe overflow, Thorne-Zytkow objects \citep{Thorne1977, Levesque2014}, and common envelopes to the chemical enrichment of the Galaxy. A notable result from their `wind group' at $Z=0.0001$ is that they find common envelopes boost all elemental yields by about 3-4 orders of magnitude $\sim 4-64\,$Myr after formation. They only report on the H, He, C, N, O, Ne, Mg, Si, S, Ca, and Fe. In our binary populations, we find an overproduction of about 4 orders of magnitude of the total C and N ejected $\sim 40-50\,$Myr after formation. However, the contribution to the total C and N from our binary population $40-50\,$Myr after formation are on the order of $10^{-9}$ and $10^{-10} \Msun/{\rm M_{\odot,SFM}}$, respectively, which is insignificant compared to the total C and N yield of our stellar populations.

There are multiple potential explanations for the discrepancy. The \textsc{binary\_c} models evolved for \citet{Yates2024} include stars with initial masses up to $120\Msun$, whereas we only include stars up to $7\Msun$. Common envelopes and Roche-lobe overflow events with massive stars will contribute to the total yield of the `wind group'. Also, their treatment of the common envelope is almost identical to ours, except for their choice to use a constant binding energy efficiency parameter $\lambda_{\rm CE} = 0.5$, whereas we use a variable $\lambda_{\rm CE}$ dependent on the stellar mass and radius as described in \citet{Dewi2000}.

\subsection{Uncertainty in Intermediate-Mass Models}
\label{Disc:IMS}
Mass-loss through stellar winds on the TP-AGB introduces significant uncertainty to the lifetimes and hence the yields of hot-bottom burning stars. Models from \citet{Doherty2014_2} and \citet{Gil-Pons2021} indicate this uncertainty increases with decreasing metallicity. Observations of intermediate-mass hot-bottom burning stars are vital to constrain our models. Unfortunately, the only stars observed with sufficient resolution at $Z=0.0001$ exist within the Galactic halo. Stars born within the Galactic halo are $\sim 10\,$Gyr in age, with only stars of mass $\lesssim 0.8\Msun$ currently surviving. Previous studies \citep{Izzard2009, Pols2012} have used observations of N-enhanced metal-poor stars to constrain their models and identify three objects with $-2.8 \leq  \text{[Fe/H]} \leq -1.8$. However, they do not consider the possibility of contamination of N-enhanced objects born in globular clusters within the Galactic halo following a merger \citep{Horta2021, Changmin2023}. Observations of white dwarfs in the Galactic halo may be another option \citep{Romero2015}; however, there are only a few known observations of massive white dwarfs ($\gtrsim 0.8\Msun$) in the Galactic halo \citep{Torres2021}. 

Since \textsc{binary\_c} models are based on fits to single stars, incorrect assumptions are likely made when evolving stars within binaries. For example, our \textsc{binary\_c} models will allow stars to extend their lifetimes on the TP-AGB and extend hot-bottom burning if they accrete additional material after evolving off the main sequence. In this scenario, stars enter the TP-AGB with a relatively low-mass core compared to their total mass, reducing their stellar radii and mass-loss rates. 

Models described in \citet{Osborn2023} show that if material is accreted during or before core He burning, the star might not evolve onto the AGB with a low-mass core. Instead, they found the core grows to mass similar as predicted for a single star of the new total mass during core He burning, and the star evolves like a single star on the AGB without any major extension of the AGB lifetime like predicted from our models. Note that the work of \citet{Osborn2023} only evolve two detailed stellar models to explore post-merger hot-bottom burning stars, they do not explore sufficient parameter space for us to implement this into \textsc{binary\_c}.

Detailed binary-star models are necessary to address the incorrect single-star assumptions applied to stellar evolution within binaries. Next-generation binary population synthesis models such as MINT \citep{Mirouh2023}, METTISE \citep{Agrawal2020}, and POSYDON \citep{Fragos2023} evolve their models based on fits to binary detailed models; however, they currently do not model AGB stellar evolution and nucleosynthesis. 
 
\subsection{Uncertainty Introduced by Binary Evolution}
Throughout this paper, we have discussed the uncertainty introduced by mass-loss on the TP-AGB, but not from binary effects such as mass transfer and common envelopes. One of the most poorly constrained binary mechanisms is the evolution of a common envelope system \citep{Ivanova2013}. Our models utilize the common envelope prescription described in \citet{Webbink1984, Tout1997}, where energy from the stellar orbit is transferred to the common envelope with an efficiency $\alpha_{\rm CE}$. $\alpha_{\rm CE}$ influences whether or not a common envelope system results in a stellar merger or the ejection of the common envelope. Many binary population synthesis codes adopt this formalism \citep{Hurley2002, Izzard2004, Riley2022, Fragos2023}. In this study, we have used the default $\alpha_{\rm CE} = 1$. However, this might not be accurate for all stellar systems \citep{Politano2004, Iaconi2019, Hirai2022}. Since the outcomes of a common envelope event have vastly different consequences on the subsequent stellar evolution and nucleosynthesis of the involved stars, it is important to quantify the impact of $\alpha_{\rm CE}$.

The TP-AGB formation rate ranges from $0.151 \pm 0.006$ per ${\rm M_{\odot, SFM}}$ for $\alpha_{\rm CE} = 0.1$ to $0.113 \pm 0.07$ per ${\rm M_{\odot, SFM}}$ for $\alpha_{\rm CE} = 5$. In the case of hot-bottom burning TP-AGB stars, the rates range from $(1.76 \pm 0.03) \times 10^{-2}$ per ${\rm M_{\odot, SFM}}$ when $\alpha_{\rm CE} = 0.1$ to $(1.14 \pm 0.03) \times 10^{-2}$ per ${\rm M_{\odot, SFM}}$ when $\alpha_{\rm CE} = 5$. When $\alpha_{\rm CE} = 0.1$, our binary populations average a $(16 \pm 1)\%$ reduction in the ejected C and a $(16 \pm 11)\%$ increase in the ejected N compared to our single star populations. When $\alpha_{\rm CE} = 5$ we find a $(47 \pm 1)\%$ reduction in the ejected C and a $(41 \pm 5)\%$ reduction in the ejected N. Our choice of $\alpha_{\rm CE}$ introduces more uncertainty to the amount of C and N ejected by our stellar population than our choice of mass-loss prescription on the TP-AGB. 

Observations of C-enhanced metal-poor stars might help constrain our treatment of $\alpha_{\rm CE}$ and binary evolution in general. However, previous studies exploring binary mechanics have shown that populations modelled using \textsc{binary\_c} do not reproduce all their observed frequencies and abundances \citep{Izzard2009, Abate2015_2}. Advancements in the treatment of binary mechanisms, such as stellar wind accretion \citep{Saladino2019}, mass transfer \citep{Temmink2023}, and common envelope evolution \citep{Gonzalez2022, Hirai2022}, may help improve our models. However, observational surveys estimate the fraction of C-enhanced metal-poor stars in the metal-poor stellar population ($\text{[Fe/H]} = -2$) to be about $10-30\%$ \citep{Lucatello2006, Lee2013, Placco2014}. Observational surveys of C-enhanced metal-poor stars do not always agree with one another due to selection effects, uncertainties, and biases in the spectral analysis \citep{Arentsen2022}, which limits our ability to reliably constrain our models. Presently, binary evolution remains a significant source of uncertainty for the chemical output of a stellar population.

\subsection{Excluding the Yield Contribution from Novae and Supernovae}
Throughout our work, we have excluded the contribution of supernovae and novae from our stellar yields. Our focus on the evolution of AGB stars motivated this choice. However, low and intermediate-mass stars are required for explosions such as Type Ia supernovae, which contribute significantly to the iron-peak elements \citep{Iwamoto1999, Kobayashi2006, Keegans2023, Cavichia2024}. Additionally, mergers and mass accretion may lead to stars born of intermediate-mass to gain sufficient material to explode in a core-collapse or electron-capture supernova. 

At a binary fraction of 0, our populations have an average electron-capture supernova rate of $(1.50 \pm 0.08)\times 10^{-3}$ and core-collapse supernova rate of up to $2  \times 10^{-4}$ per ${\rm M_{\odot, SFM}}$. These originate from stars of initial mass between about $6.2-7\Msun$, with $7\Msun$ being the maximum initial mass we model in our stellar populations. At a binary fraction of 1, the average electron capture supernova rate decreases to $(1.09 \pm 0.02) \times 10^{-3}$ per ${\rm M_{\odot, SFM}}$ and the core-collapse supernova rate increases to $(2.0 \pm 0.1) \times 10^{-3}$ per ${\rm M_{\odot, SFM}}$. We also find an average type 1a supernova rate of $(1.48 \pm 0.05) \times 10^{-4}$ per ${\rm M_{\odot, SFM}}$ from our binary star populations. However, we do not construct our models with the goal of measuring supernova rates and therefore, do not consider these rates reliable. 

Previous studies have explored the rates of novae \citep{Kemp2022} and supernovae \citep{Ruiter2011, Zapartas2017}, including their yield contribution \citep{Izzard2003, Izzard2004_Thesis, Yates2024, Kemp2024}. The omission of novae and supernovae will likely not introduce significant uncertainty to the yields of key elements such as C, N, F, and s-process elements due to the dominance of production within AGB stars \citep{Kobayashi2020}, but given sufficient frequency they will impact elements such as O, Na, Mg, Al, and the iron peak elements.

\subsection{Comparing the results of our Delay Time Distributions to Observed Populations}

Our delay-time distributions in Figure \ref{fig:CN_Delay} show, for example, our binary populations overproduce N by over an order of magnitude during the period $\sim 300-700\,$Myr after formation, compared to our single star populations. To verify these results and those for the other elements we study, we need to compare our predictions with the abundances observed in stellar populations.

Predictions from galactic chemical evolution models are mostly compared to the abundances of unevolved low-mass field stars \citep{DeDonder2002, Valentini2019, Kobayashi2020, Molero2025}. A similar comparison using the ejecta from our models, however, is not informative. The surface compositions of field stars in the Galactic halo \citep{Fulbright2002, Venn2004} are mostly representative of their abundances at birth, and their stellar ages are not well enough defined to disentangle the individual generations of stars. For each simulated population, we do not attempt to calculate how the ejected material mixes with the material in the interstellar medium, nor do we calculate the composition of the following generation of stars.

A comparison to metal-poor globular clusters would also not be informative. Their ages are relatively well resolved \citep{Valcin2020}, and multiple stellar populations can be identified \citep{Ziliotto2023, Howell2024}. However, globular cluster stars are chemically anomalous compared to Galactic stars of the same metallicity \citep{Hendricks2014, Gratton2019}. They also have such high stellar densities that dynamical interactions become significant, resulting in their current-day binary fraction to be $\lesssim 10\%$ \citep{Ivanova2005}.

In its current state, we are unable to directly compare our calculated delay time distributions to observed stellar populations, such as the Galactic Halo. Our delay time distributions are not designed to infer the ages of a given observed stellar population. They are designed to provide an estimate of how elements such as N can be expelled into the Galaxy as a function of time, owing to stellar and binary evolution. The most informative step would be to use our yields within a Galactic chemical evolution code and evolve the abundances as a function of time, for comparison to field stars in different stellar populations. That work, however, is beyond the scope of this study.

\section{CONCLUSION}
\label{sec:Conclusion}
We have used the binary population synthesis code \textsc{binary\_c} to stellar populations from five model sets with various mass-loss prescriptions on the TP-AGB at $Z=0.0001$. We have found that for our populations with a binary fraction of 1, the formation rate of TP-AGB stars reduces by about $37\%$ compared to our populations calculated with a binary fraction of 0. This correlates with our binary populations ejecting about $38\%$ less C and about $35-40\%$ less s-process elements than our single-star populations. Our binary populations also produce about $32\%$ fewer hot-bottom burning stars. Our choice of mass-loss prescription introduces significant uncertainty to the chemical output of our hot-bottom burning models. However, we find an overproduction of N over an order of magnitude in our binary star population $\sim 300-700\,$Myr after formation. The role of wind uncertainty is far less significant on our lower mass stars ($\lesssim 3\Msun$).

Binary evolution adds significant uncertainty to our models. Our treatment of common envelope evolution varies the formation rate of TP-AGB stars in our binary population from about $0.113$ per ${\rm M_{\odot, SFM}}$ to $0.151$ per ${\rm M_{\odot, SFM}}$, introducing a significant variation to the C and N yields. Future work will refine the treatment of mass transfer and common envelope events in our models and explore how they influence the population yields in detail. 

\begin{acknowledgement}

ZO acknowledges this research was supported by an Australian Government Research Training Program (RTP) Scholarship. DK acknowledges funding support from the Australian Research Council Discovery Project DP240101150. AJK acknowledges financial support from the Flemish Government under the long-term structural Methusalem funding program by means of the project SOUL: Stellar evolution in full glory, grant METH/24/012 at KU Leuven. RGI is funded by STFC grants ST/Y002350/1, ST/L003910/1, and ST/R000603/1 as part of the BRIDGCE UK network. CK acknowledges funding from the UK Science and Technology Facilities Council through grant ST/Y001443/1.

This work was performed on the OzSTAR national facility at Swinburne University of Technology. The OzSTAR program receives funding in part from the Astronomy National Collaborative Research Infrastructure Strategy (NCRIS) allocation provided by the Australian Government, and from the Victorian Higher Education State Investment Fund (VHESIF) provided by the Victorian Government.

We used the package from \url{https://github.com/keflavich/imf} to generate the initial mass distribution of our single and binary primary stars.

We thank the anonymous referee for their detailed suggestions, which helped us to improve our paper.

For the purpose of Open Access, the author has applied a Creative Commons Attribution (CC BY) public copyright licence to any Author Accepted Manuscript version arising from this submission.

\end{acknowledgement}

\paragraph{Data Availability Statement}
Tables presenting the total and net yields of all populations and the delay-time distributions of C, N, F, Sr, Ba, and Pb will be made available online upon publication. Additional data can be made available upon reasonable request to the corresponding authors. 

The official release of \textsc{binary\_c} version 2.2.4 is available at \url{https://gitlab.com/binary_c/binary_c/-/tree/releases/2.2.4?ref_type=heads}.


\bibliography{bibtemplate}


\end{document}